\documentclass[aps,prb,twocolumn,showpacs,amsmath,amssymb]{revtex4}
\usepackage{epsfig}
\usepackage{graphicx}
\begin{document}
\newcommand{\bn}{{\bf n}}
\newcommand{\bp}{{\bf p}}   
\newcommand{\br}{{\bf r}}
\newcommand{\bH}{{\bf H_0}}
\newcommand{\eps}{\varepsilon}
\newcommand{\tz}{\hat\tau_z}
\newcommand{\ty}{\hat\tau_y}
\newcommand{\cG}{\check G}
\newcommand{\hG}{\hat G}
\newcommand{\la}{\langle}
\newcommand{\ra}{\rangle}
\newcommand{\ua}{{\mathbf\uparrow}}
\newcommand{\da}{{\mathbf\downarrow}}
\newcommand{\be}{\begin{equation}}
\newcommand{\ee}{\end{equation}}
\newcommand{\fs}{f_{\sigma}}
\newcommand{\fms}{f_{-\sigma}}
\newcommand{\ns}{n_{\sigma}}
\newcommand{\nms}{n_{-\sigma}}
\newcommand{\Ss}{\sum\limits_{\sigma=\pm 1}}
\newcommand{\Sa}{\sum\limits_{\alpha}}
\newcommand{\gs}{\gamma_{\scriptscriptstyle\Sigma}}
\newcommand{\Ses}{S_{\sigma}^i}

\newcommand{\mybeginwide}{
    \end{multicols}\widetext
    \vspace*{-0.2truein}\noindent
    \hrulefill\hspace*{3.6truein}
}
\newcommand{\myendwide}{
    \hspace*{3.6truein}\noindent\hrulefill
    \begin{multicols}{2}\narrowtext\noindent
}

\title{Current fluctuations in a spin filter with paramagnetic impurities  }

\author{ K. E. Nagaev$^{1,2}$ and L. I. Glazman$^1$ }

\affiliation{$^1$W. I. Fine Theoretical Physics Institute, University of Minnesota, Minneapolis, Minnesota 55455}

\affiliation{
$^2$Institute of Radioengineering and Electronics, Russian Academy of Sciences, Mokhovaya ulica 11, 125009 Moscow, Russia}

\date\today

\begin{abstract}
  We analyze the frequency dependence of  shot noise in a spin filter
  consisting of a normal grain and ferromagnetic electrodes separated
  by tunnel barriers. The source of frequency-dependent noise is
  random spin-flip electron scattering that results from spin-orbit
  interaction and magnetic impurities. Though the  latter mechanism
  does not contribute to the average current, it contributes to the
  noise and leads to its dispersion at frequencies of the order of the
  Korringa relaxation rate.  Under nonequilibrium conditions,
  this rate is proportional to the applied bias $V$, but parametrically smaller than $eV/\hbar$.
\end{abstract}


\maketitle

Recently, fundamental and applied aspects of spin-dependent transport
attracted significant attention of physicists.\cite{Zutic-04} One of the
prototype devices realizing this type of transport is a spin filter
based on the use of ferromagnetic (FM) electrodes. In such a device, one of the electrodes
may be viewed as "polarizer" that generates a spin-polarized current
and the other, the "analyzer", collects it. Current through the
spin filter can drastically vary depending on the mutual orientation
of electrode magnetizations.

An important characteristic of a mesoscopic device is its
nonequilibrium noise,\cite{Blanter-00} which is due to randomness of
electron transport and provides an important information about its
nature.  Fluctuations of current in a quantum dot with FM electrodes were considered in
the Coulomb blockade regime\cite{Bulka-99,Bulka-00} and without it.\cite{Gurvitz-05} Nonequilibrium noise in a diffusive
conductor with different mutual orientations of magnetizations in
the leads was also studied.\cite{Tserkovnyak-01} More recently,
several authors focused on the noise in diffusive spin filters caused
by spin-flip scattering, which plays a crucial role in transport
between electrodes with different magnetizations. This work was
pioneered by Mishchenko,\cite{Mishchenko-03} who analyzed the noise in
two-terminal diffusive spin filters for parallel and antiparallel
magnetizations of electrodes and found a significant increase of shot
noise in the latter case as compared to the conventional diffusive
contacts. Calculations of shot noise caused by spin-flip scattering
were also performed by other authors.\cite{Lamacraft-04} Shot noise and
cross-correlations of current  for multiterminal
spin filters with spin-flip scattering were calculated\cite{Belzig-04,Zareyan-05} as well. A
frequency dependence of the shot noise in a quantum dot with spin-flip
scattering connected with FM electrodes by ballistic point
contacts was calculated\cite{Mishchenko-04} by Mishchenko {\it et al.}

In the absence of a magnetic field, spin-flip scattering in metals is
dominated by spin-orbit processes and collisions of electrons with
magnetic impurities. Though both these processes affect the spin of an
individual electron and manifest themselves in weak-localization
corrections to conductivity, only spin-orbit scattering contributes to
the relaxation of average spin current; scattering off magnetic
impurities conserves the total spin of the electron -- impurity system
and therefore the impurities eventually become polarized by incident
electrons. The goal of the present paper is to study the relative
importance of these two scattering mechanisms for the shot noise in spin filters.
The principal difference between them is that the spin--orbit
scattering changes only the spin of the electron and not the state of
the ordinary impurity from which it is scattered. In contrast to this,
spin-flip scattering off magnetic impurities results in a change of
both impurity and electron spins thus affecting the ability of an
impurity to flip an electron spin in the future. This results in
impurity-induced correlations between electrons and therefore the
spectral density of the noise of current should reflect their
dynamics. Typically, the electron spin-orbit
scattering time is much shorter than the time of spin-exchange scattering on paramagnetic impurities.\cite{Pierre-03} Therefore, the impurity spin-relaxation time should
manifest itself in the frequency dependence of the noise spectrum. We
calculate the spectrum of the noise in a spin filter at frequencies
much smaller than the charge-relaxation time and show that it exhibits
features characteristic of the impurity-spin relaxation.

\section{Equations for the averages }

The system we consider consists of a normal-metal grain connected with
two ferromagnetic electrodes via tunnel junctions with conductions
much smaller than that of the grain yet larger than $e^2/\hbar$ so that 
Coulomb blockade does not occur.
The level spacing is negligible compared to charging energy
$e^2/C$ in a typical metallic grain. We also neglect the electron--phonon and
electron--electron scattering. The polarizations of
electrodes lead to spin-dependent tunneling rates, as the latter are
proportional to the density of states of electrons with a given spin
projection.

The average distribution functions of spin-up and spin-down electrons,
$\fs$ $(\sigma = \pm 1)$ are described by equations
\begin{eqnarray}
 \frac{\partial\fs}{\partial t}
 &=&
 \Gamma_{L\sigma}[f_0(\eps - eV/2) - \fs]
\nonumber\\
 &+&
 \Gamma_{R\sigma}[f_0(\eps + eV/2) - \fs]
\nonumber\\
 &-&
 \frac{1}{\tau_{so}} (\fs - \fms) 
 +
 \Ses 
 -
 e \frac{\partial U}{\partial t} 
 \frac{\partial\fs}{\partial\eps}.
 \label{f-eq}
\end{eqnarray}
Here $\Gamma$'s are the tunneling rates through the left and right
barriers for spin-up and spin-down electrons, $f_0(\eps)$ is the
equilibrium distribution function of electrons, $\tau_{so}$ is the
spin-orbit scattering time, and $\Ses(\eps)$ is the electron--impurity
collision integral. The spin-flip part of  the collision integral,
evaluated in the lowest (second) order of perturbation theory in $J$,
has the form

\begin{equation}
 \Ses(\eps)
 =
 J^2 N_F
 [\ns\fms (1 - \fs) - \nms \fs (1 - \fms)].
 \label{S_e}
\end{equation}
where $J$ is the constant of exchange interaction between the itinerant
electrons and impurity spin, $N_F$ is the density of electron levels
in the grain, $\ns$ are concentrations of spin-up and
spin-down impurities, and $U$ is the electric potential of the grain.
All scattering processes accounted for in Eq.~(\ref{f-eq})  are elastic (in the second order in $J$, relaxation
of electron energy in scattering off magnetic impurities does not
occur\cite{Kaminski-01}). 

The relaxation of impurity spins in metals is also mainly due to
interaction with electrons. In equilibrium, this mechanism is known
as Korringa relaxation\cite{Korringa-50} and the corresponding
relaxation rate is proportional to the electron temperature. In the
case of an arbitrary electron distribution, the dynamics of impurities
is described by a kinetic equation
\begin{equation}
 \frac{\partial\ns}{\partial t}
 =
 -N_F\int d\eps\, \Ses.
 \label{n1-eq}
\end{equation}
This equation is compatible with the conservation law $\Ss\ns = n \equiv const$.  The stationary
solutions of these equations are found from the conditions
$$
 \frac{\partial\fs}{\partial t}=0,
 \qquad
 \frac{\partial U}{\partial t}=0,
 \qquad
 \Ses = 0.
$$
It suggests that under the stationary conditions, the magnetic
impurities do not affect the average distribution function of
electrons and current. However they still contribute to the noise, as
will be shown below.

\section{ The Langevin equations }

The dynamics of fluctuations is conveniently described by a set of
coupled Langevin equations for the relevant distribution functions.
Such equations are derived by varying properly the kinetic equations,
and adding Langevin sources to the result of variations.\cite{Kogan} For the spin filter we consider,
the proper distribution functions are those of spin-up and spin-down
electrons and magnetic impurities. The Langevin equations for these
functions read
\begin{eqnarray}
 &\Biggl(&
   \frac{\partial}{\partial t}
   +
   \Gamma_{L\sigma} + \Gamma_{R\sigma} + \frac{1}{\tau_{so}}
   -
   \frac{\partial\Ses}{\partial\fs}
  \Biggr)
  \delta\fs
\nonumber\\%$$\vspace{-3mm}$$
  &=&
  \left(
   \frac{1}{\tau_{so}}
   +
   \frac{\partial\Ses}{\partial\fms}
  \right)
  \delta\fms
  +
  \frac{\partial\Ses}{\partial\ns}\delta\ns
  +
  \frac{\partial\Ses}{\partial\nms}\delta\nms
  %\sum_{\sigma'=\pm 1}\frac{\partial\Ses}{\partial n_{\sigma'}}\delta n_{\sigma'}
\nonumber\\%$$ \vspace{-3mm}\be
  &-&
  e\frac{\partial\fs}{\partial\eps}
  \frac{\partial\delta U}{\partial t}
  -
  \delta j_{L\sigma} - \delta j_{R\sigma} 
  + 
  \delta F_\sigma^{so} + \delta F_\sigma^i,
  \label{Lf1}
 \end{eqnarray}
where $\Ses$ is given by Eq. (\ref{S_e})
and $\delta j_{\alpha\sigma}$, $\delta F_\sigma^{so}=\sigma F_{so}$, and $\delta F_\sigma^i=\sigma F_i$ are
Langevin sources related to tunneling of electrons with spin $\sigma/2$
through the left ($\alpha=L$) and right
($\alpha=R$) barriers, spin-orbit scattering, and spin-impurity
scattering, respectively. The dynamics of the impurities is described
by  Langevin equations
\begin{eqnarray}
  &\Biggl[&
   \frac{\partial}{\partial t}
   +
   N_F \int d\eps\, 
   \left(
     \frac{\partial\Ses}{\partial\ns}
     -
     \frac{\partial\Ses}{\partial\nms}
    \right)
  \Biggr]
  \delta\ns
\nonumber\\% $$\vspace{-3mm}\be
  &=&
  -N_F\int d\eps\,
  \left(
   \frac{\partial\Ses}{\partial\fs}
   \delta\fs
   +
   \frac{\partial\Ses}{\partial\fms}
   \delta\fms
  \right)
\nonumber\\
  &-&
  N_F \int d\eps\, \delta F_\sigma^i,
  \label{Ln1}
 \end{eqnarray}
where the relation $\delta\ns = -\delta\nms$ was taken into account.
 
 Fluctuations of the currents flowing into the left and right
 electrodes can be expressed in terms of $\delta f$ and $\delta j$,
\begin{equation}
 \delta I_\alpha 
 =
 e N_F \sum_{\sigma=\pm 1}\int d\eps\,
 ( \Gamma_{\alpha\sigma}\delta\fs + \delta j_{\alpha\sigma} ),
 \qquad
  \alpha=L,R. \label{dI}
\end{equation}
To complete the set of equations, we relate the fluctuation of the
electric potential of the grain $\delta U$, to the fluctuations of the
currents flowing into the grain. Such relation follows from the charge
conservation law and the definition of the electrostatic capacity of
the grain $C$:
\begin{equation}
 \frac{\partial}{\partial t}
 \delta U
 =
 -\frac{1}{C}
 ( \delta I_L + \delta I_R ).
 \label{dU}
\end{equation}

It is natural to treat the Langevin sources $\delta j_{\alpha\sigma}$,
$\delta F_{so}$, and $\delta F_{imp}$ as independent because they
correspond to different scattering processes.  As the scattering is
assumed to be weak, it may be considered Poissonian, and the
correlation functions of the Langevin sources in these equations may
be written as the sums of outgoing and incoming scattering fluxes,
\begin{eqnarray} 
 \la\delta F_{so}(\eps_1, t_1) \delta F_{so}(\eps_2, t_2)\ra 
 =
 \delta(t_1 - t_2) \delta(\eps_1 - \eps_2)
\nonumber\\%$$\vspace{-3mm}\be
 \times
 \frac{1}{N_F \tau_{so}}
 \Ss
 \fs(\eps_1) [1 - \fms(\eps_1)],
 \label{<Fso^2>}
\end{eqnarray}
\begin{eqnarray}
 \la\delta F_{imp}(\eps_1, t_1) \delta F_{imp}(\eps_2, t_2)\ra 
 =
 \delta(t_1 - t_2) \delta(\eps_1 - \eps_2)
\nonumber\\%$$\vspace{-5mm}\be
 \times
 \frac{J^2}{N_F}
 \Ss
 \ns \fms(\eps_1) [1 - \fs(\eps_1)],
 \label{<Fi^2>}
\end{eqnarray}
\begin{eqnarray}
 \la\delta j_{\alpha\sigma}(\eps_1, t_1) \delta j_{\alpha\sigma}(\eps_2, t_2)\ra
 =
 \delta(t_1 - t_2) \delta(\eps_1 - \eps_2)
\nonumber\\%$$\vspace{-5mm}\be
 \times
 \frac{\Gamma_{\alpha\sigma}}{N_F}
 \Bigl\{
  \fs(\eps_1) [1 - f_0(\eps_1 - eV/2)]
\nonumber\\
  +
  f_0(\eps_1 - eV/2) [1 - \fs(\eps_1)]
 \Bigr\}.
 \label{<j1^2>}
\end{eqnarray}
Equations (\ref{Lf1}) - (\ref{<j1^2>}) allow us to find the correlation functions of any observable quantity, e.g. of the current.

\section{ Parallel magnetization of electrodes }

First consider the case of a parallel magnetization of electrodes.
Suppose that both left and right electrodes have spin-up magnetization
so that only spin-up electrons can cross the interfaces with the
normal metal. This can be modeled by setting
$\Gamma_{L,-1}=\Gamma_{R,-1}=0$. The stationary solution of Eqs.
(\ref{f-eq}) and (\ref{n1-eq}) is
$$
 \fs(\eps)
 =
 \frac
 {\Gamma_{L1} f_0(\eps - eV/2) + \Gamma_{R1} f_0(\eps + eV/2) }
 {\Gamma_{L1} + \Gamma_{R1}},
$$\vspace{-5mm}\be
 \ns = n/2.
 \label{avs-p}
\end{equation}
As we are interested in frequencies much smaller than the inverse
charge-relaxation time, we may set $C=0$ in Eq. (\ref{dU}), which
reduces it to $\delta I_L + \delta I_R =0$. Together with Eqs.
(\ref{dI}), this gives
\begin{equation}
 \delta I_L
 =
 e N_F \int d\eps\,
 \frac
 { \Gamma_{R1}\delta j_{L1} + \Gamma_{L1}\delta j_{R1} }
 { \Gamma_{L1} + \Gamma_{R1} }.
 \label{dI_L-p}
\end{equation}
Making use of Eqs. (\ref{<j1^2>}) and (\ref{avs-p}),
one readily obtains that in the low-temperature limit,
\begin{equation}
 \la\delta I_L^2\ra_{\omega}
 =
 e^3 N_F
 \frac
 { \Gamma_{L1}\Gamma_{R1} (\Gamma_{L1}^2 + \Gamma_{R1}^2) }
 {(\Gamma_{L1} + \Gamma_{R1})^3}
 V,
 \label{<I^2>-p}
\end{equation}
while the average current is
\begin{equation}
 I = e^2 N_F \frac{\Gamma_{L1}\Gamma_{R1}}{\Gamma_{L1} + \Gamma_{R1}} V.
 \label{<I>-p}
\end{equation}
These results suggest that for the parallel magnetizations of electrodes, neither the average current nor the noise are affected by spin-flip processes and the whole system behaves like a normal double tunnel junction with the density of states reduced by half. This is in a marked contrast with the case of a normal grain whose material has nonzero  resistance,\cite{Mishchenko-03} where the spin-down sub-band in the grain contributes substantially to the current.

\section{ Antiparallel magnetization of electrodes }

In the case of antiparallel magnetization of the leads, electron
transport is associated with the spin-flip processes in the
normal grain (we assume that the magnetization in the leads is
saturated, and $\Gamma_{L,-1} = \Gamma_{R1} = 0$). Considering for
simplicity the case of symmetric junctions, $\Gamma_{L1} = \Gamma_{R,-1} =
\Gamma$, we find
\begin{equation}
 \fs
 =
 \frac{
  (1 + \Gamma\tau_{so}) f_0(\eps - \sigma eV/2)
  +
  f_0(\eps + \sigma eV/2)
 }{
  2 + \Gamma\tau_{so}
 }.
 \label{<f1>}
\end{equation}
The steady impurity concentrations $\ns$  are
determined by setting $\partial\ns/\partial t=0$ in
Eq.~(\ref{n1-eq}), and accounting for the conservation condition $\ns + \nms=n$,
\begin{equation}
 \frac{\ns}{n}
 =
 \frac
 { \int d\eps\, \fs(1 - \fms) }
 { \int d\eps\, \Ss \fs(1 - \fms)}.
 \label{<n1n2>}
\end{equation}

Our goal now is to obtain closed Langevin equations for fluctuations $\delta\ns$ and $\delta U$ by eliminating $\delta\fs$ from them. Then we can substitute the solutions of these equations back into the equation for $\delta\fs$ and calculate the fluctuation of current.

Fluctuations of the distribution functions $\delta\fs$ of the
spin-up  and spin-down electrons can be found from the
Langevin equations (\ref{Lf1})
\begin{eqnarray}
 \delta\fs(\eps, \omega)
 &=&
 \frac{-1}{-i\omega + \gs(\eps)}
 \Biggl\{
   \frac{1}{-i\omega + \Gamma}
   \Biggl[
    (-i\omega + \Gamma + \eta_{-\sigma})   
\nonumber\\
   &\times&
    \left(
     -i\omega  
     \frac{\partial\fs}{\partial\eps} 
     e\delta U
     +
     \Sa\delta j_{\alpha\sigma}
    \right)
\nonumber\\
    &+&
    \eta_{-\sigma}
    \Biggl(
     -i\omega
     \frac{\partial\fms}{\partial\eps}
     e\delta U  
     +
     \Sa\delta j_{\alpha,-\sigma}
    \Biggr)
   \Biggr]   
\nonumber\\%$$ \vspace{-5mm} \be
   &-&
   \frac{\partial\Ses}{\partial\ns}
   \delta\ns
   -
   \frac{\partial\Ses}{\partial\nms}
   \delta\nms
   +
   \delta F_\sigma^{so} + \delta F_\sigma^i
 \Biggl\},
   \label{df1-sol}
\end{eqnarray}
where
$$
 \eta_\sigma
 =
 \frac{1}{\tau_{so}}
 -
 \frac{\partial\Ses}{\partial\fs},
$$
and
$$
 \gs(\eps) = \Gamma + \eta_\sigma + \eta_{-\sigma}
$$
is the effective relaxation rate due both to tunneling and all types of spin-flip scattering.
It follows from Eq.~(\ref{<f1>}) that $1 - \fms(\eps) = \fs(-\eps)$,
and therefore
\be
 \left.
  \frac{ \partial\Ses }{ \partial\fs }
 \right|_{\eps}
 =
 \left.
  \frac{ \partial S_{-\sigma}^i }{ \partial\fms }
 \right|_{-\eps}.
 \label{deriv}
\ee
Hence, $\eta_\sigma(\eps) = \eta_{-\sigma}(-\eps)$ and $\gs(\eps) = \gs(-\eps)$. 

By substituting $\delta\fs$  from Eq.~(\ref{df1-sol})
into Eq.~(\ref{Ln1}), one obtains for the fluctuations of impurity polarization
\begin{eqnarray}
 \delta\ns
 &=&
 \frac{\sigma N_F}{ -i\omega + \Omega(\omega) }
 \int d\eps\,
 \Biggl[ 
  (X - 1) \delta F_{imp} + X \delta F_{so} 
\nonumber\\%$$\vspace{-5mm}\be
  &+& 
  \Sa\Ss \sigma Y_\sigma(\eps, \omega) \delta j_{\alpha\sigma}
 \Biggr].
 \label{dn1-3}
\end{eqnarray}
Here
\begin{equation}
 \Omega(\omega)
 =
 N_F \int d\eps\,
 \frac
 { -i\omega + \Gamma + 2/\tau_{so} }
 { -i\omega + \gs(\eps) }
 \frac{\partial S_1^i}{\partial n_1}
 \label{Omega}
\end{equation}
gives at $\omega\to 0$ the inverse relaxation time of the impurity spins;
coefficients
\begin{equation}
 X(\eps, \omega)
 =
 -\frac{1}{ -i\omega + \gs(\eps) }
 \Ss 
 \frac{\partial\Ses}{\partial\fs},
 \label{X}
\end{equation}
and
\begin{equation}
 Y_\sigma(\eps, \omega)
 =
 \frac
 { 
   [1 + \tau_{so}(-i\omega + \Gamma)]
   \partial\Ses/\partial\fs
   +
   \partial\Ses/\partial\fms
 }
 {
   \tau_{so} ( -i\omega + \Gamma )[ -i\omega + \gs(\eps) ]
 }
 \label{Y1}
\end{equation}
relate the fluctuations $\delta\ns$ to Langevin sources.
Note that $Y_\sigma(\eps, \omega)=Y_{-\sigma}(-\eps, \omega)$. 

Using Eqs. (\ref{dI}) and (\ref{dU}) we now calculate  $\delta U$
\begin{equation}
 i\omega C \delta U
 =
 eN_F \int d\eps\, \Ss
 \left(
  \Gamma\delta\fs + \Sa\delta j_{\alpha\sigma}
 \right).
%[\Gamma ( \delta f_1 + \delta f_2 ) + \delta j_{L1} + \delta j_{R2}].
 \label{dU2}
\end{equation}
Substituting the sum of solutions from Eq. (\ref{df1-sol})
$$
 \Ss\delta\fs
 =
 -\frac{1}{-i\omega + \Gamma}
 \Ss
 \left(
  -i\omega
  \frac{\partial\fs}{\partial\eps}
  e\delta U
  +
  \Sa\delta j_{\alpha\sigma}
 \right)
$$
into Eq. (\ref{dU2}) and integrating out the derivatives, one obtains a closed expression for $\delta U$ in the form\cite{Nagaev-04}
\begin{equation}
 \left(
  -i\omega + \Gamma + 2\frac{e^2 N_F\Gamma}{C}
 \right)
 \delta U
 =
 -\frac{eN_F}{C}
 \int d\eps\,
 \Sa\Ss \delta j_{\alpha\sigma}.
 \label{dU-expl}
\end{equation}
Because the grain is sufficiently large, the second term in the parentheses in the left-hand side of Eq. (\ref{dU-expl}) may be neglected as compared to the third term. The large coefficient $e^2 N_F/C$ in the last term in parentheses is proportional to the ratio of the charging energy to the level spacing.
As we are interested in frequencies much smaller than the inverse charge-relaxation time, Eq. (\ref{dU-expl}) reduces to
\begin{equation}
 \delta U
 =
 -\frac{1}{e\Gamma}
 \int d\eps\,
 \Sa\Ss \delta j_{\alpha\sigma}.
%(\delta j_{L1} + \delta j_{R2}).
 \label{dU-lf}
\end{equation}

Now we are able to express the fluctuation of current $\delta I_L$ in terms of Langevin sources $\delta J_{\alpha\sigma}$, $\delta F_{imp}$, and $\delta F_{so}$.
For that, we substitute now Eq. (\ref{df1-sol}) for $\delta\fs$ into Eq. (\ref{dI}) for $\delta I_L$. Making use of the identity (\ref{deriv}),
one may eliminate the integration over $\eps$.
Substitution of $\delta U$ from Eq. (\ref{dU-lf}) and $\delta n_1$ from Eq. (\ref{dn1-3}) into the resulting expression gives
\begin{eqnarray}
\delta I_L
 &=&
 e N_F \int d\eps\,
 \Biggl\{
  \Ss
  \Biggl[
   \sigma
   \frac
   { 
    -i(\omega/2)
    (-i\omega + \gs)
    +
    \Gamma\eta_\sigma
   }{
    (-i\omega + \Gamma)
    (-i\omega + \gs)
   } 
\nonumber\\
   &-&
   \frac
   { \Gamma\Omega Y_\sigma(\eps,\omega)}
   { (-i\omega + \Gamma + 2/\tau_{so})(-i\omega + \Omega) }
  \Biggr]
  \Sa\delta j_{\alpha\sigma}
\nonumber\\
  &+&
  \Gamma
  \Biggl[
   \frac
   {1}
   {-i\omega + \gs}
   +
   \frac
   {\Omega(1 - X)}
   { (-i\omega + \Gamma + 2/\tau_{so})(-i\omega + \Omega) }
  \Biggr]
\nonumber\\
  &\times&
  \delta F_{imp}
  +
  \Gamma
  \Biggl[
   \frac
   {1}
   {-i\omega + \gs}
\qquad\nonumber\\
   &-&
   \frac
   {\Omega X}
   { (-i\omega + \Gamma + 2/\tau_{so})(-i\omega + \Omega) }
  \Biggr]
  \delta F_{so}
 \Biggl\}.
 \label{dI-2}
 \end{eqnarray}
As the Langevin sources $\delta j_{\alpha\sigma}$,  $\delta F_{imp}$, and $\delta F_{so}$ are independent, multiplying Eq. (\ref{dI-2}) by its complex conjugate and taking into account the expressions for the correlators of these sources (\ref{<Fso^2>}) - (\ref{<j1^2>}), one obtains the spectral density of the noise as the sum of three terms related with the three types of scattering
\begin{equation}
 \la\delta I_L^2\ra_{\omega}
 =
 \Ss P_\sigma(\omega) + P_{so}(\omega) + P_i(\omega),
 \label{I^2}
\end{equation}
where
\begin{eqnarray}
 P_\sigma(\omega)
 &=&
 e^2 N_F \Gamma \int d\eps\,
 \Biggl|
  \frac
  {
   -i(\omega/2)
   (-i\omega + \gs)
   +
   \Gamma\eta_\sigma
  }{
   (-i\omega + \Gamma)
   (-i\omega + \gs)
  }
\nonumber\\%$$\vspace{-3mm}$$
  &-&
  \frac
  {
   \sigma\Gamma \Omega(\omega) Y_\sigma(\eps, \omega)
  }{
   (-i\omega + \Gamma + 2/\tau_{so})
   [-i\omega + \Omega(\omega)]
  }
 \Biggr|^2
\nonumber\\%$$  \vspace{-5mm}\begin{equation}
 &\times&
 \Bigl\{
  \fs(\eps) [1 - f_0(\eps - \sigma eV/2)]
\nonumber\\
  &+&
  f_0(\eps - \sigma eV/2) [1 - \fs(\eps)]
 \Bigr\},
 \quad\label{P1}
\end{eqnarray}
\begin{eqnarray}
 P_{so}(\omega)
 &=&
 \frac{e^2 N_F \Gamma^2}{\tau_{so}} \int d\eps\,
 \Biggl|
  \frac
  {1}
  {-i\omega + \gs}
\nonumber\\
  &-&
  \frac
  { 
   \Omega(\omega)X(\omega) 
  }{ 
  (-i\omega + \Gamma + 2/\tau_{so})
  (-i\omega + \Omega)  
  }
 \Biggr|^2
\nonumber\\
 &\times&
 \Ss \fs(1 - \fms),
 %\Bigl[f_1(1 - f_2) + f_2(1 - f_1)\Bigr]
 \label{Pso}
\end{eqnarray}
and
\begin{eqnarray}
 P_i(\omega)
 &=&
 e^2 N_F J^2 \Gamma^2 \int d\eps\,
 \Biggl|
  \frac
  {1}
  {-i\omega + \gs}
\nonumber\\
  &+&
  \frac
  { 
   \Omega(\omega)[1 - X(\omega)] 
  }{ 
  (-i\omega + \Gamma + 2/\tau_{so})
  (-i\omega + \Omega)  
  }
 \Biggr|^2
\nonumber\\
 &\times&
 \Ss n_{\sigma} \fms (1 - \fs).
 %\Bigl[n_1 f_2(1 - f_1) + n_2 f_1(1 - f_2) \Bigr].
 \label{Pi}
\end{eqnarray}

The general expressions Eqs. (\ref{I^2}) - (\ref{Pi}) for the spectral density are too cumbersome to be analyzed,  and therefore we restrict ourselves to several interesting limiting cases. We assume that in all cases $T=0$.

The average current through the system is
\begin{equation}
 I = \frac{ e^2 N_F\Gamma }{ 2 + \Gamma\tau_{so} } V.
 \label{<I>}
\end{equation}
Note that $I \to 0$ at $\Gamma\tau_{so} \to \infty$ as no electron can pass through the filter without flipping its spin.

The noise properties of a system are often described by a frequency-dependent Fano factor $F=\la\delta I_L^2\ra_{\omega}/I$. In the absence of magnetic impurities, the Fano factor is
\begin{equation}
 F
 =
 \frac{1}{2} e
 \frac
 {
  \tau_{so}^2
  (\omega^2 + 2\Gamma^2)
  (2 + \Gamma\tau_{so})
  +
  8 (1 + \Gamma\tau_{so})
 }{ 
  (2 + \Gamma\tau_{so})
  [ \omega^2\tau_{so}^2 + (2 + \Gamma\tau_{so})^2]
 }.
 \label{no_imp}
\end{equation}
It exhibits positive dispersion for small $\Gamma\tau_{so}$ and  negative dispersion for large $\Gamma\tau_{so}$ (see Fig. 1).
%\ref{fig:nimp}). 
Its dependence on the dwell time of an electron $1/\Gamma$ in the grain with tunnel junctions  is different from the case of a grain with ballistic contacts\cite{Mishchenko-04} both at zero and nonzero frequencies.

The presence of paramagnetic impurities leads to an additional frequency dispersion of the noise. The characteristic frequency $\omega_0$ of that dispersion is of the order of inverse spin-flip time for these impurities. At $T=0$ and within the lowest order perturbation theory in $J$, it is proportional\cite{omega0} to the applied bias $V$ and therefore can be made much smaller than $\tau_{so}^{-1}$. In what follows, we will assume that this is the case. As the noise now depends on $V$ nonlinearly, it is more convenient to consider its full value rather than the Fano factor. In the limit $\Gamma\tau_{so} \ll 1$,
\begin{eqnarray}
 \Delta\la\delta I_L^2\ra_{\omega}
 &\equiv&
 \la\delta I_L^2\ra_{\omega}
 -
 \la\delta I_L^2\ra_{\infty}
\nonumber\\
 &=&
 \frac{1}{8}
 e^2 n (\Gamma\tau_{so})^2
 \frac
 {
   \omega_0^3(V) 
 }{
  \omega^2 + \omega_0^2(V)
 },
 \label{short_tau}
\end{eqnarray}
where $\omega_0(V) = eVN_F^2 J^2/2$ is an analogue of inverse Korringa relaxation time with $eV$ in place of $T$.
The low-frequency dispersion originates from the fluctuations of the number of spin-up and spin-down impurities, which modulate the overall spin-flip rate of electrons in the grain and result in fluctuations of the transport current. It is absent in the case of purely spin-orbit scattering.\cite{Mishchenko-04}

For arbitrary values of $\Gamma\tau_{so}$, one can make use of the difference in the frequency scales between the spin-orbit and impurity-spin relaxations and isolate the part of 
$\la\delta I_L^2\ra_{\omega}$ that exhibits low-frequency dispersion. It has a form similar to Eq. (\ref{short_tau})
\begin{eqnarray}
 \Delta\la\delta I_L^2\ra_{\omega}
 =
 %2 e^2 N_F J^2 (eV)
 e^2 n 
 A(\Gamma\tau_{so})
 \frac{\Omega^3(0)}{\omega ^2 + \Omega^2(0)}, 
 \label{I^2-lf}
\end{eqnarray}
with a modified  spin-flip frequency of impurities
\begin{eqnarray}
 \Omega(0)
 =
 2\omega_0(V)
 B(\Gamma\tau_{so}),
\qquad%\nonumber\\
 B(x)
 =
 \frac{ 1 + (1 + x)^2 }{ (2 + x)^2 }
 \label{Omega-lf}
\end{eqnarray}
and the dimensionless amplitude $A$ given by
\begin{equation}
A(x)
 =
 2
 \frac
 { 
  x^2
  (1 + x)
  ( 2 + 5x + x^2 )
 }{
  (2 + x)^3
  [1 + (1 + x)^2]^2
 }.
 \label{A}
\end{equation}
\begin{figure}[t]
 \includegraphics[width=8.5cm]{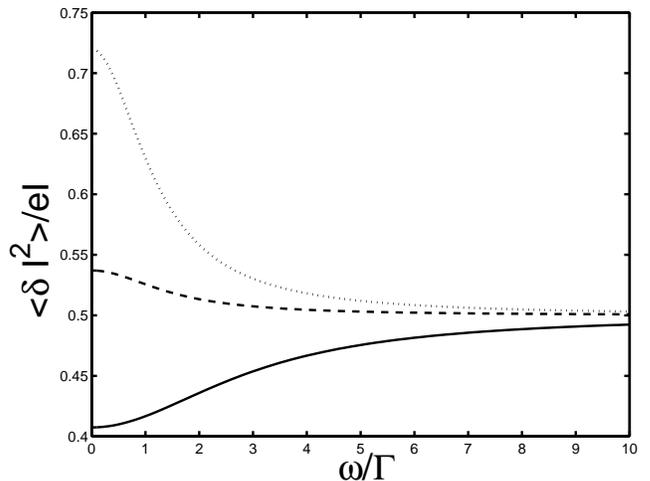}
 \caption{\label{fig:nimp} The frequency dependence of the Fano factor in the absence of magnetic impurities for  
 $\Gamma\tau_{so}=1$ (solid line), $\Gamma\tau_{so}=4$ (dashed line), and $\Gamma\tau_{so}=10$ (dotted line).}
\end{figure}
\begin{figure}[t]
 \includegraphics[width=8.5cm]{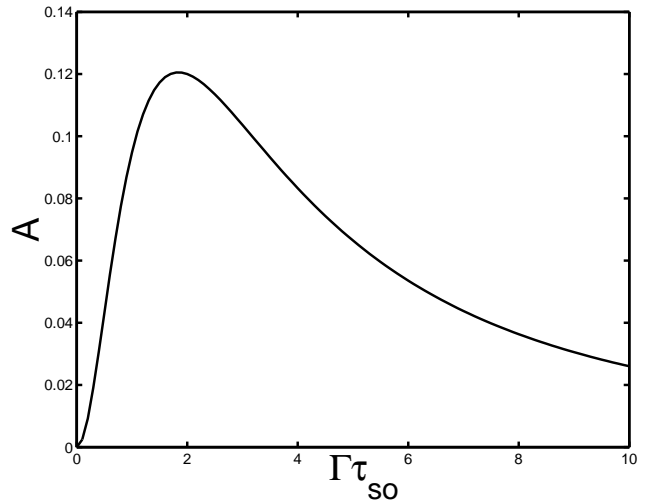}
 \caption{\label{fig:amplot} The amplitude of $\Delta\la\delta I_L^2\ra_{\omega}$ normalized to $e^2n\omega_0(V)$ 
 vs. $\Gamma\tau_{so}$.
 } 
\end{figure}
As $\Gamma\tau_{so}$ increases from 0 to infinity, the factor $B$ monotonically increases from 1/2 to 1.
The dimensionless amplitude $A$ %of this part  normalized to $e^2n\omega_0(V)$ 
is plotted in Fig. 2 %\ref{fig:amplot} 
as a function of $\Gamma\tau_{so}$. It vanishes both at $\Gamma\tau_{so}\ll 1$ because in this case the two spin orientations become equivalent and the system behaves essentially as a one-sub-band conductor, and at $\Gamma\tau_{so}\gg 1$ because in this case the impurity spins become completely polarized and no spin-flip transitions take place.

\section{ Summary }

In summary, we have shown that paramagnetic impurities contribute to the shot noise in ferromagnet - normal metal - ferromagnet spin filters with opposite magnetizations of electrodes. Though their contribution to the noise is smaller than the contribution of spin-orbit scattering, it can be distinguished by a characteristic low-frequency dispersion that results from impurity-spin reorientations. As the rate of the impurity-spin relaxation depends on the energy distribution of electrons in the normal metal, this dispersion is affected by the applied voltage. Though the present calculations were performed for the ideal case of completely polarized electrodes, these effects also take place for more realistic systems with partial polarization.

\begin{acknowledgments}

This work was supported by NSF grants DMR 02-37296, DMR 04-39026 and EIA 02-10736.

\end{acknowledgments}


\begin{thebibliography}{99}
\bibitem{omega0}
It implies that  bias $eV$ is much larger than the Kondo temperature of the magnetic impurities in the given host metal.
\end{thebibliography}
\end{document}